\documentstyle[prb,aps,twocolumn,tighten,psfig,epsf,floats]{revtex}
\begin{document}
\draft
\title{Correlated theory of triplet photoinduced absorption in 
phenylene-vinylene chains}
\author{Alok Shukla}
\address{Physics Department, Indian Institute of Technology, Bombay,
Powai, Mumbai 400076 INDIA} 
\maketitle
\begin{abstract}
 In this paper we present  results of large-scale correlated 
calculations of  triplet photoinduced absorption (PA) spectrum of 
oligomers of poly-(para)phenylenevinylene (PPV) containing up to five 
phenyl rings.  In particular, the high-energy features in the triplet PA
spectrum of oligo-PPVs are the focus of this study, which, so far,  have 
not been investigated theoretically, or experimentally. The calculations were 
performed using the Pariser-Parr-Pople (PPP) model Hamiltonian,  
and many-body effects were taken into account by means of 
multi-reference singles-doubles configuration
interaction procedure (MRSDCI), without neglecting any molecular orbitals. 
The computed triplet PA spectrum of oligo-PPVs exhibits rich structure 
consisting of alternating peaks of high and low intensities. 
The predicted higher energy features of the triplet spectrum can be tested 
in future experiments.  Additionally, theoretical estimates of exciton 
binding energy are also presented. 
\end{abstract}
\section{Introduction}
 Photoluminescent conjugated polymer poly-(para)phenylenevinylene (PPV) is
one of the most promising candidates for the
optoelectronic devices of the next generation, such as light-emitting
diodes,~\cite{friend} and lasers.~\cite{lasers}
Because of  their quasi one-dimensional  character, the electrons(e) and 
holes(h) in conjugated polymers have a high recombination probability, 
leading to their large photoluminescence (PL) efficiency.~\cite{friend}
However, in electroluminescence experiments, a significant fraction of 
the e-h recombination events lead to the formation of triplet excitons, 
which are nonluminescent.~\cite{wohl} In addition, it is believed that through 
various other processes such as intersystem crossing,~\cite{oster} and 
fission,~\cite{cerullo} the singlet excitons 
in these materials transform into the
triplet ones. Therefore, in order to fully characterize the PL properties
of conjugated polymers such as PPV, a thorough understanding of their triplet 
excited states
is required. A good understanding of the triplet states is fruitful from
another vantage point as well. The energy difference between the triplet 
and singlet states defines the exchange energy, and consequently, quantifies
the strength of e-e correlations in the system. Therefore,
a deeper understanding of the  triplet states will
also enhance our understanding of electron correlation effects 
in conjugated polymers.
Recently \"Osterbacka et al.,~\cite{oster} Romanovskii et al.,~\cite{bassler}
and Monkman et al.~\cite{monkman} have measured the
energies of the first singlet ($S_1$) and triplet ($T_1$)
excited states of various conjugated polymers. 
Triplet excited states with energies higher than $T_1$ are frequently
probed through the photoinduced absorption (PA) spectroscopy in the triplet 
manifold.~\cite{cerullo,oster} In these
experiments first the $T_1$ is populated, and then the optical
transitions are induced to higher-energy triplet states.  
Assuming a planar configuration for oligo-PPVs corresponding to symmetry
group $C_{2h}$, their one- and two-photon states will belong, respectively, 
to the $B_u$, and the $A_g$ irreducible representations. The ground state of 
these oligomers is $1^1A_g$, while $S_1$ corresponds to $1^1B_u$. 
In the triplet manifold,  $T_1$ corresponds to the $1^3B_u$ state, while the 
first state with a strong dipole coupling to $1^3B_u$, is called $m^3A_g$. Therefore, in the
triplet PA spectrum of oligo-PPVs, $m^3A_g$ would show up as the
first peak, while any higher $^3A_g$-type states with significant 
dipole coupling to $1^3B_u$, would lead to high energy features in the 
spectrum. Therefore, through triplet PA spectroscopy,
one can probe the triplet two-photon states ($^3A_g$) of oligo-PPVs. However,
so far all the theoretical,~\cite{chandross,beljonne,barford} and 
experimental~\cite{oster,cerullo} investigations of triplet PA in PPVs 
have focused on the first $m^3A_g$ state, providing no information on
the higher-energy $^3A_g$-type states. Considering recent PA measurements
of high-energy two-photon states in the singlet manifold ($^1A_g$) of the 
PPVs which have yielded a lot of exciting information on the nature of those
states~\cite{frolov}, we believe that a similar analysis should be extended
to the high-energy $^3A_g$-type states of this material.
Therefore, to fill that that void, we decided to undertake a systematic 
large-scale correlated calculation of the triplet PA spectrum of 
oligo-PPVs which will target
the high-energy features at the same level of computational accuracy as the
$m^3A_g$ peak. However, compared to the singlet-$A_g$ manifold responsible 
for the singlet PA , the correlation effects are stronger in the
triplet-$A_g$ manifold because here,
in addition to the orbital excitations, spin excitations are also involved.
Therefore, it is mandatory that the correlated calculations employ large
basis sets in order to describe the high-energy $^3A_g$ peaks accurately.
Since these are the first calculations of the kind, for the
purpose of benchmarking we also performed calculations on a number of
other excited states of oligo-PPVs, and compare them
to the works of other authors. 
As a matter of fact our calculations on the triplet PA spectrum indicate the
existence of three peaks beyond $m^3A_g$  exhibiting rich structure with 
alternating high and low intensities. 
We hope that this work will encourage experimental measurements of the
high-energy region in the triplet PA spectrum of PPVs, just like 
similar recent measurements of their singlet PA spectrum.~\cite{frolov}

The remainder of this paper is organized as follows. In section \ref{method}
we briefly describe the correlation approaches used to perform the 
calculations in the present work. Next in 
section \ref{results} we present and discuss the calculated triplet PA spectrum
of oligo-PPVs. Finally, in section \ref{conclusion} we summarize our 
conclusions, and discuss possible directions for future work.

\section{Methodology}   
\label{method}

Henceforth any oligomer of PPV containing $n$ phenyl rings will be 
referred to as PPV-$n$, making {\em trans}-stilbene synonymous with
PPV-2. We performed correlated calculations on the oligomers PPV-2
to PPV-5, assuming completely planar geometries and the symmetry group
 $C_{2h}$. From among a variety of model Hamiltonians
available,~\cite{beljonne} we chose the Pariser-Parr-Pople (PPP) model for
the present calculations, which describes the $\pi$-electron dynamics
of conjugated systems in terms of a minimal basis set, and a 
small number of parameters.
The  PPP Hamiltonian reads
\begin{eqnarray}
\label{H_PPP}
H  & = & - \sum_{\langle ij \rangle, \sigma} t_{ij}
(c_{i \sigma}^\dagger c_{j\sigma}+ c_{j \sigma}^\dagger c_{i \sigma}) +
U \sum_i n_{i \uparrow} n_{i \downarrow} \nonumber \\
  &  & +   \sum_{i<j} V_{ij} (n_i -1)(n_j -1)
\end{eqnarray}
\noindent where $\langle ij \rangle$ implies nearest neighbors,
$c_{i \sigma}^\dagger$ creates an electron of spin
$\sigma$ on the $p_z$
orbital of carbon atom $i$, $n_{i \sigma} = c_{i \sigma}^\dagger c_{i \sigma}$
is the number of electrons with spin $\sigma$, and
$n_i = \sum_\sigma n_{i \sigma}$ is the total number of electrons on atom $i$.
The parameters $U$ and $V_{ij}$ are the on--site
and long--range Coulomb interactions, respectively, while $t_{ij}$ is the
nearest-neighbor one-electron hopping matrix element.
The Coulomb interactions are parametrized as per Ohno relationship, \cite{ohno}
\begin{equation}
V_{i,j} = U/\{\kappa_{i,j} (1+0.6117R_{i,j}^2)^{1/2} \} \; \mbox{,}
\label{eq-ohno}
\end{equation}
where $R_{i,j}$ is the distance (in $\AA$) between the sites $i$ and $j$,
and $\kappa_{i,j}$ is the dielectric constant to account for screening.
Since the results obtained with the PPP Hamiltonian depend substantially
on the choice of the Coulomb parameters, we tried two sets: (a)
``standard parameters'' with
 $U=11.13$ eV and $\kappa_{i,j}=1.0$, and (b) ``screened
parameters'' with $U=8.0$ eV and $\kappa_{i,i}=1.0$, and $\kappa_{i,j} = 2.0$
with $i\neq j$. The screened parameters, originally introduced by
Chandross and Mazumdar~\cite{chandross} in their study of optical absorption
in PPV, are known to give better agreement with experiments on excitation
energies, possibly by taking interchain screening effects into account.  
For the nearest-neighbor hopping matrix elements $t_{ij}$, 
we chose $-2.4$ eV for the phenylene ring, and $-2.2$ eV and $-2.6$ eV,
respectively, for the 
single and double bonds of the vinylene linkage. As far as the bond lengths
are concerned, in the phenyl rings they were taken to be 1.4 $\AA$. In the
vinylene linkage the single (double) bond lengths were taken to be 
1.54 $\AA$ (1.33 $\AA$). 

Starting with the restricted Hartree-Fock (HF), depending on the
size of the oligomer, the electron correlation effects
were included either by the quadruple configuration-interaction (QCI) method or
by multi-reference singles-doubles configuration-interaction (MRSDCI) 
approach. Both the approaches are well documented in the 
literature.~\cite{tavan}
The accuracy of the MRSDCI calculations depends on the choice of the reference
 configurations, and their number ($N_{ref}$). The reference 
configurations are chosen according to their contribution to the
many-particle wave function of the targeted state, while the
convergence of our results with respect to $N_{ref}$ will be examined in the
next section. 
To reduce the size of the Hamiltonian 
matrices, we made full use of the spin and the $C_{2h}$ point-group symmetry.
The triplet  PA spectrum was calculated using the oscillator strengths
computed under the electric dipole approximation. The finite lifetime of
the states was taken into account by means of  a linewidth parameter $\Gamma$.
Further technical details can be found in our recent work, where we used 
this symmetry-adapted  CI methodology to study the low-lying excited states in
phenyl-substituted polyacetylenes.~\cite{pdpa-prb} 

\section{Results and Discussion}
\label{results}
\begin{figure}
\begin{center}
\psfig{file=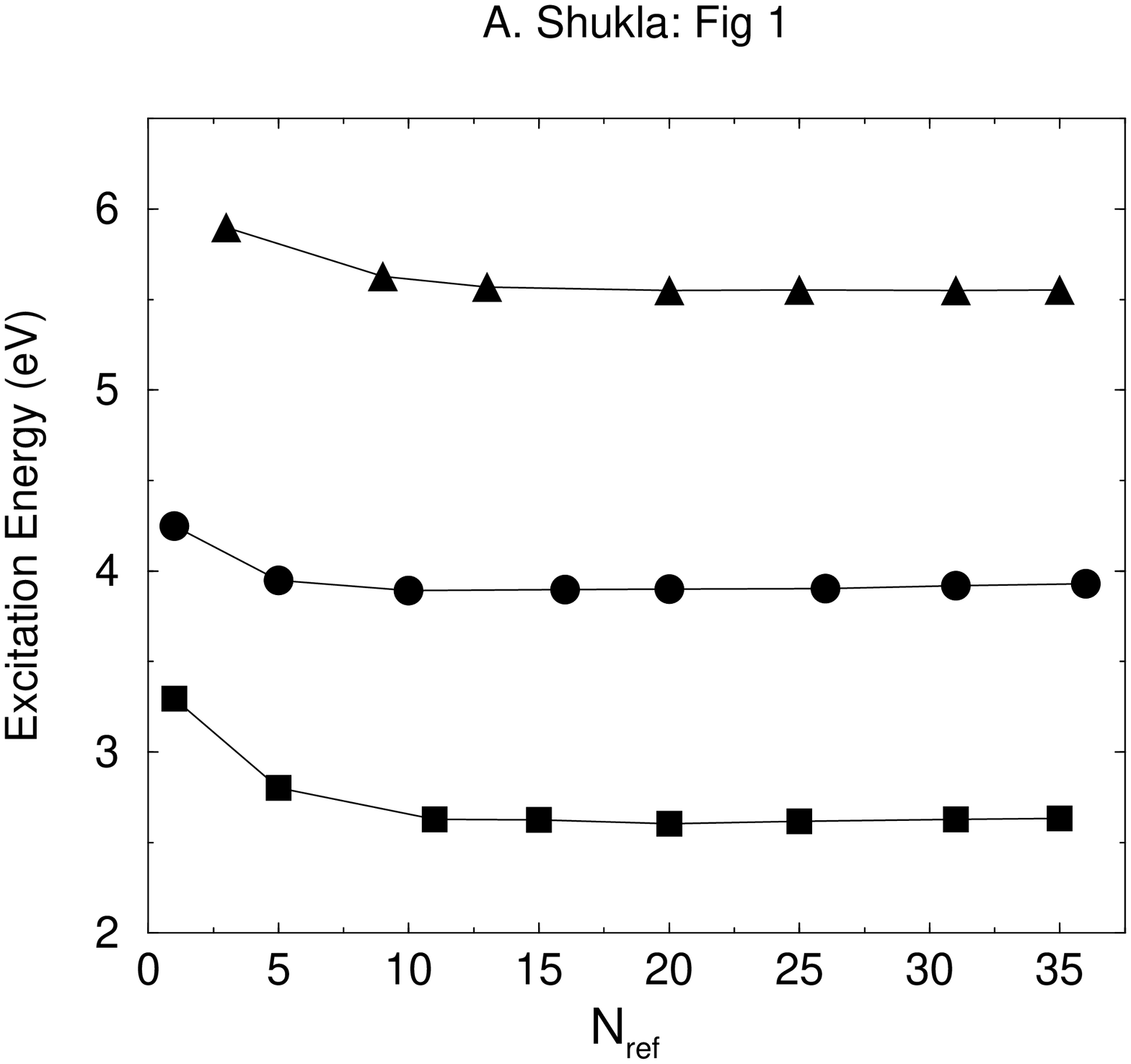,width=8.0cm,angle=-90}
\end{center}
\caption{Behavior of $1^3B_u$ (squares), $1^1B_u$ (circles), and
$m^3A_g$ (triangles) excited states  of PPV-4 with respect to the
number of reference configurations ($N_{ref}$) included in the
MRSDCI calculations performed with standard parameters.}
\label{fig-conv}
\end{figure}

The correlation method, $N_{ref}$, and the total number of configurations
($N_{total}$) involved in the CI calculations of excited states of 
various oligo-PPVs are detailed in table \ref{tab-nref}.
From the sizes of the CI matrices diagonalized  it is clear that these
calculations were very large scale, and therefore, they must include the
influence of electron correlation effects to a very high accuracy.
The convergence of our MRSDCI results with respect to $N_{ref}$, is
demonstrated in Fig. \ref{fig-conv}.  It is clear
from the results that good convergence in the excited state energies
has been achieved in all the cases by the time fifteen most important
configurations ($N_{ref}=15$) have been included in the reference space
in various MRSDCI calculations, leading us to believe that the
excitation energies presented here are accurate
to within a few hundredths of an eV, for the given values of PPP parameters. 

The noteworthy aspect of the
present calculations as compared to those of other 
authors~\cite{beljonne,gartstein,barford} is that no
MOs have been discarded from the correlated calculations. 
Lavrentiev et al.,~\cite{barford} with the purpose of obtaining results
for infinite PPV, performed density-matrix 
renormalization group calculations in conjunction with a 
PPP-Hamiltonian-based two-state model consisting of just the highest-occupied 
and the lowest unoccupied MOs (HOMO and LUMO) of each phenylene and vinylene
unit.  Thus, e.g.,  for PPV-5 which has 38 MOs in total, the number of 
active MOs considered by Lavrentiev et al.~\cite{barford} was just 18. 
Additionally,   Lavrentiev et al.~\cite{barford} used a different set of
PPP parameters as compared to us. 
Similarly Gartstein et
al.~\cite{gartstein} simplified the PPP model to obtain analytical results
for the energy of $1^3B_u$ state, but only for infinite PPV. Rohlfing 
et al.~\cite{louie} performed first-principles calculations on both the
singlet and triplet states of infinite PPV.
Beljonne et al.,~\cite{beljonne} performed calculations on the same oligomers
of PPV as considered by us, utilizing an intermediate neglect of differential 
overlap (INDO) model Hamiltonian, also within the framework of the MRSDCI 
procedure. However, they used only four reference configurations
($N_{ref}=4$), and considered single and double excitations from a
set of five highest occupied $\pi$ orbitals into the five lowest unoccupied
$\pi^{*}$ orbitals.~\cite{beljonne} Thus, $N_{ref}$ as well as the 
number of active MOs used by Beljonne et al.~\cite{beljonne} is much smaller 
compared to our calculations. Therefore, to benchmark our computational
approach and the choice of PPP Coulomb parameters, we compare our MRSDCI
results on various low-lying excited states of PPV-$n$ ($n=2\mbox{---}5$)
to those of other authors and experiments, in table \ref{tab-comp}.

Inspection of table \ref{tab-comp} reveals the following trends:
(i) Our MRSDCI excitation energies computed  with the screened Coulomb 
parameters are lower, and in better agreement with the experiments, 
than those computed with the standard parameters. 
(ii) For $E(1^3B_u)$ our calculations predict complete saturation,
in excellent qualitative agreement with the highly-localized 
picture of the $1^3B_u$ state verified recently in the experiment  of 
\"Osterbacka et al.,~\cite{oster} reporting that $1^3B_u$ was a Frenkel 
exciton, with a radius of just $3.2$ $\AA$. However, neither our calculations,
nor those of other authors
are close to the experimental value of $E(1^3B_u) \approx 1.5$ eV for the
infinite PPV.~\cite{oster,monkman} 
(iii) As far as comparison with other authors is concerned, our screened
parameter values of $E(1^1B_u)$
are in similar agreement with the experiments as those of
Beljonne et al.,~\cite{beljonne} while they are somewhat higher than
those of Lavrentiev et al.~\cite{barford} whose   $E(1^1B_u)$ for PPV-5
is lower than the experimental value. However, our screened parameter values
of $E(m^3A_g) - E(1^3B_u)$ are consistently in much better agreement with
experiments as compared to those of Beljonne et al.~\cite{beljonne} and
Lavrentiev et al.~\cite{barford} This is very encouraging because, in order
to compute the triplet PA spectrum accurately, one needs to have an accurate
representation of states in the $^3A_g$ manifold, starting with the
$m^3A_g$ state.

To further benchmark our approach, we use $E(m^3A_g)$ computed above to 
obtain a theoretical estimate
of the binding energy of the $1^1B_u$ exciton.
 Shimoi and Mazumdar have argued that, because $m^3A_g$ state is also an
 exciton, $E(m^3A_g)- E(1^1B_u)$  is a good lower-limit estimate, 
$E_b(\mbox{min})$,
for the exciton binding energy.~\cite{shimoi} 
From our calculations, for PPV-5 one gets 
$E_b(\mbox{min}) = 0.68 (1.7)$ eV with screened (standard) parameters.
 Keeping in mind that generally our screened parameter based calculations were
in good agreement with the experiments, 
we take the value $E_b(\mbox{min}) =0.68$ eV to be the correct theoretical
estimate. Thus our theoretical estimate of $E_b(\mbox{min})=0.68$ eV 
obtained from 
PPV-5  is in  very good agreement with the experimental value of 
$E_b(\mbox{min})=0.55$ eV estimated by 
\"Osterbacka et al.~\cite{oster} for infinite PPV.

\begin{figure}
\begin{center}
\psfig{file=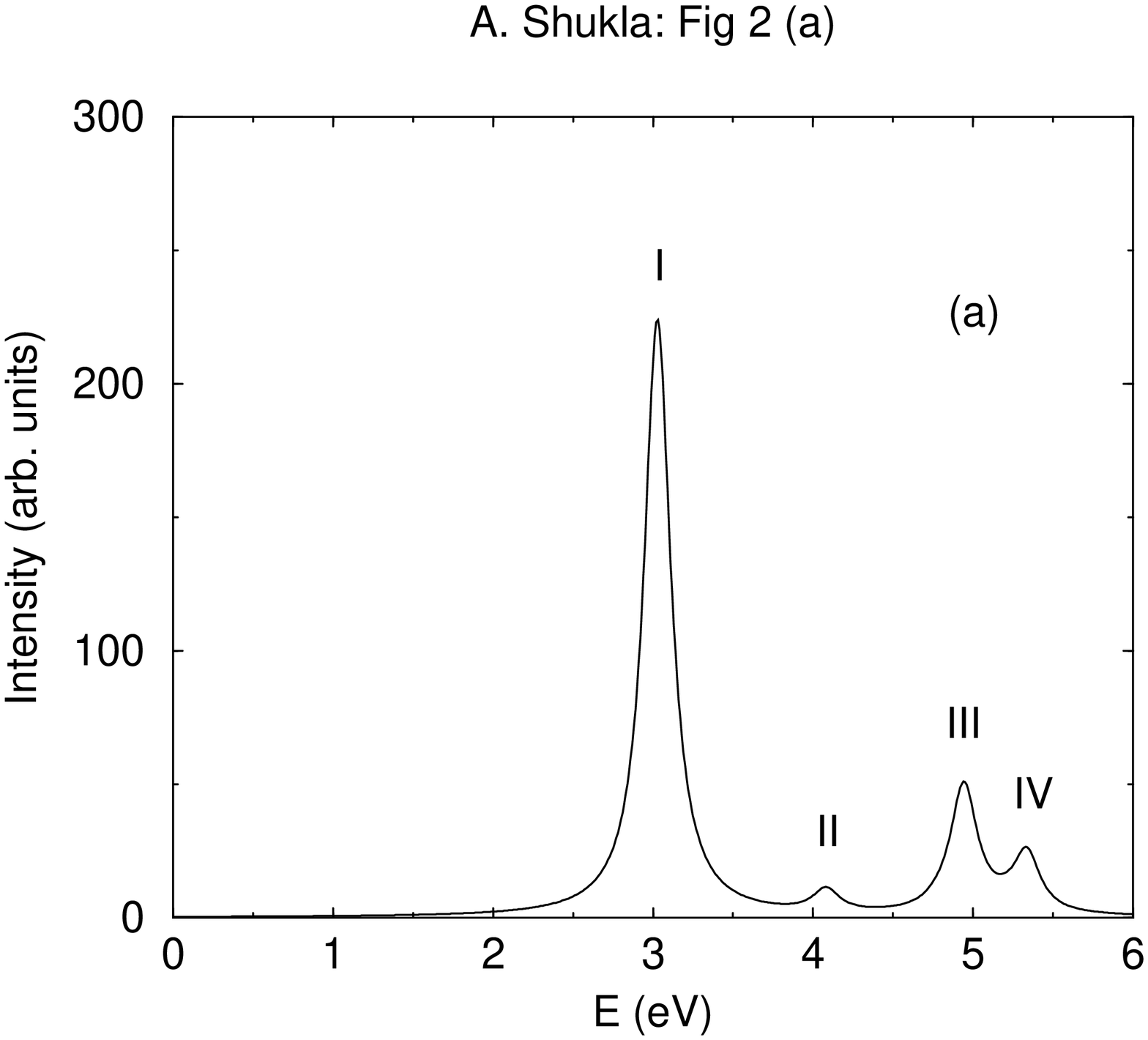,width=8.0cm,angle=-90}
\psfig{file=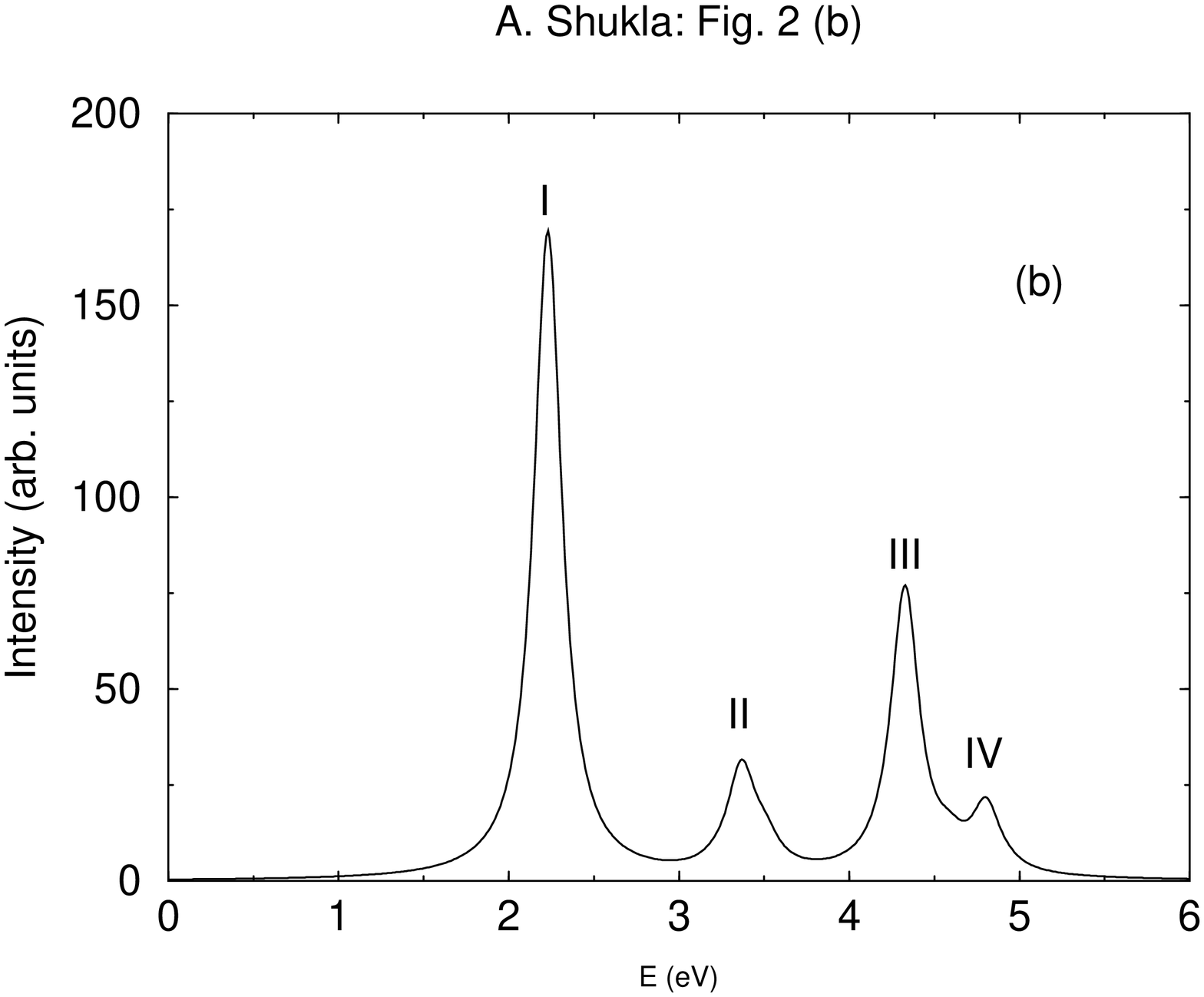,width=8.0cm,angle=-90}
\end{center}
\caption{Photoinduced absorption spectrum from $1^3B_u$ state in PPV-4:
(a) computed with the standard parameters (b) computed with the screened parameters. A linewidth parameter 
$\Gamma =0.1$ eV was used. $E$ is the energy of the incident photon.}
\label{fig-ppv4}
\end{figure}

\begin{figure}
\begin{center}
\psfig{file=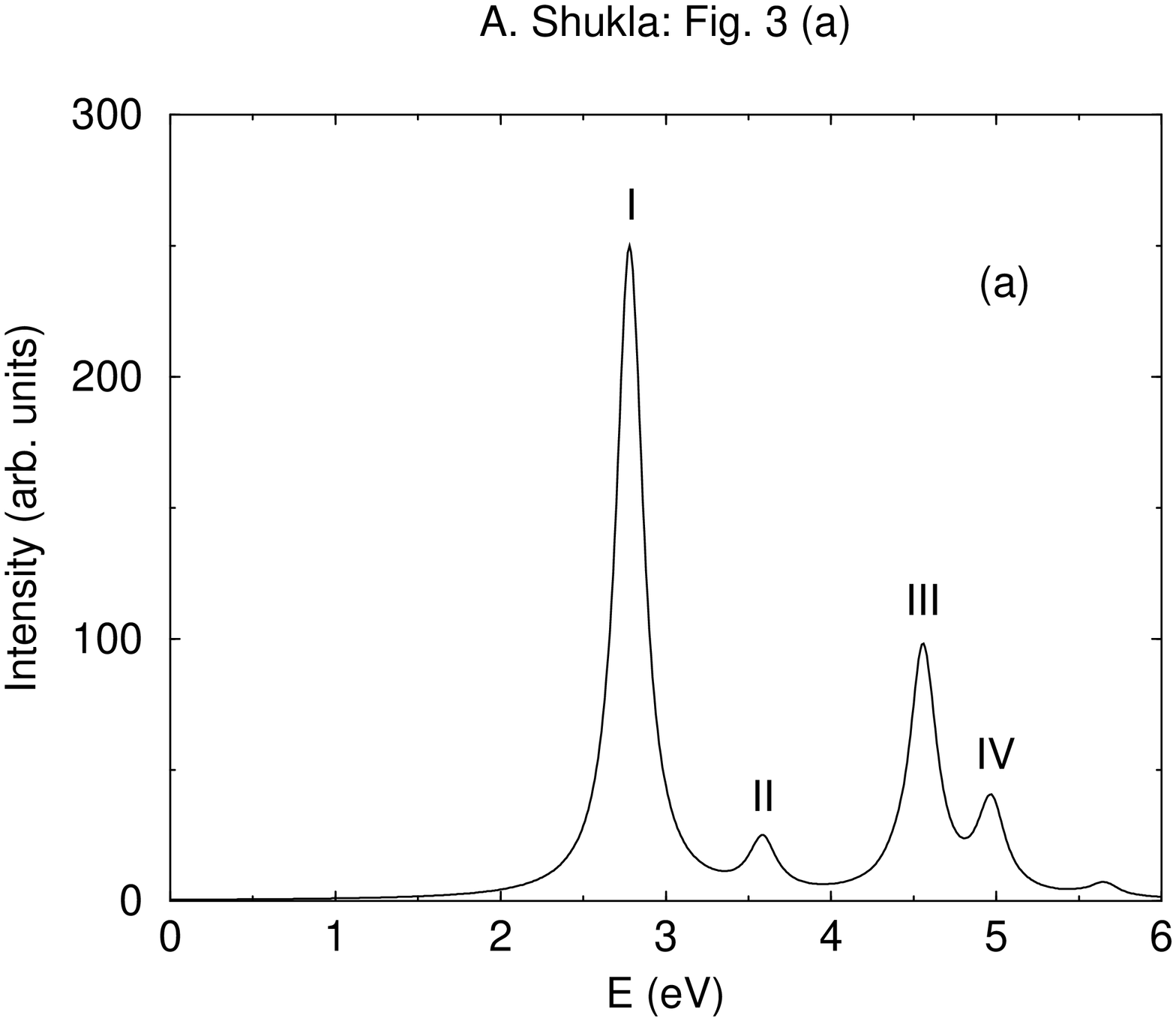,width=8.0cm,angle=-90}
\psfig{file=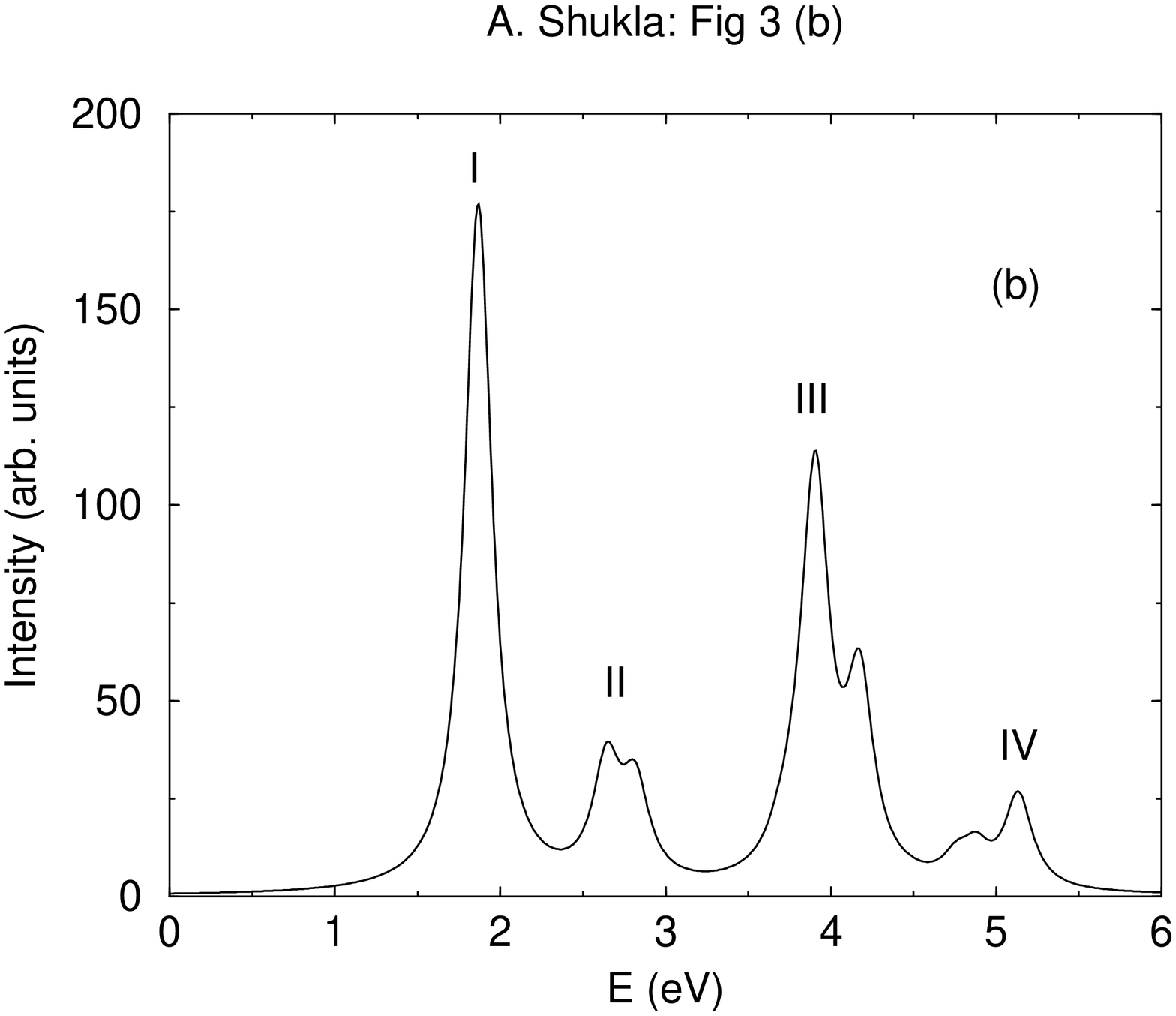,width=8.0cm,angle=-90}
\end{center}
\caption{Photoinduced absorption spectrum from $1^3B_u$ state in PPV-5:
(a) computed with the standard parameters (b) computed with the screened parameters. A linewidth parameter 
$\Gamma =0.1$ eV was used. $E$ is the energy of the incident photon.}
\label{fig-ppv5}
\end{figure}

Next we turn our attention to the main focus of this work---the triplet PA 
spectrum of oligo-PPVs. The calculated spectrum of PPV-4 is presented in
Fig. \ref{fig-ppv4}, and of PPV-5 in Fig. \ref{fig-ppv5}. The spectra
have been computed both with the standard parameters 
(Figs. \ref{fig-ppv4}a and \ref{fig-ppv5}a),  as well as with
the screened parameters (Figs. \ref{fig-ppv4}b and \ref{fig-ppv5}b).
Thus the four presented spectra will help us understand the influence of
the following two factors on the triplet PA: (i) the
conjugation length  (ii) the strength of Coulomb interactions.

The following trends emerge from the spectra presented in 
Figs. \ref{fig-ppv4} and \ref{fig-ppv5}: (i) Irrespective of the Coulomb
parameters used, both for PPV-4 and PPV-5 there are
four major structures (labeled I, II, III, and IV) of alternating high and 
low intensities. (ii) Irrespective of the Coulomb parameters, the relative 
intensities of the high energy peaks (II, III, and IV), as compared to peak 
I, are increasing with increasing conjugation length (iii) For a given
oligomer, the relative intensities of the high-energy peaks, compared to
peak I, are higher when computed using
screened parameters, as compared to the ones calculated using the standard
parameters. Although, standard-parameter-based calculations yield rather
small relative intensities for peaks II of PPV-4 and PPV-5, however,
the same parameters lead to significant intensities of other high-energy peaks 
(III and IV). Thus, based upon these results, we arrive at the following
conclusions regarding triplet PA in oligo-PPVs: (i) The intensities of
peak III and beyond are strong enough to be detected experimentally in rather
small oligomers such as PPV-4 and PPV-5. (ii) The intensity of peak II for 
small oligomers appears to be
strongly dependent on the strength of Coulomb interactions. This peak will be
detectable for these oligomers only if screened parameters, perhaps due to
interchain screening, describe the strength of Coulomb interactions. (iii) For
relatively longer oligomers, even peak II will acquire significant 
intensity both for standard and screened parameters, as exemplified by
the rapid increase in its intensity with size.  

Having discussed the general qualitative features of the computed triplet PA
spectrum, we next examine the nature of many-particle $^3A_g$ states 
contributing to various peaks. Here we limit our discussion to the
specific case of the spectrum of PPV-5 computed with screened parameters as
presented in Fig. \ref{fig-ppv5}b. The reasons behind this choice are 
obvious---PPV-5 is the longest oligomer studied, and the 
values of the energy gaps $E(m^3A_g)- E(1^1B_u)$ with screened parameters 
for various oligomers were in very good agreement with the experiments.
However, we did examine the nature of the peaks in the triplet spectra
presented in the remaining three figures (Figs. \ref{fig-ppv4}a, 
\ref{fig-ppv4}b, and \ref{fig-ppv5}a), and all of them were 
found to be qualitatively very similar to the states corresponding to
the peaks in Fig. \ref{fig-ppv5}b.  Thus we believe that our discussion of triplet
PA spectrum of PPV-5, computed with
screened parameters, represents the general features of the triplet PA spectra
of oligo-PPVs.
The first peak in Fig. \ref{fig-ppv5}b, occurring close to 1.9 eV, is 
the strongest in the spectrum, 
and corresponds to the $1^3B_u \rightarrow m^3A_g$ absorption. 
Compared to the Hartree-Fock configuration, the many-particle
wave function of $m^3A_g$ is a linear combination consisting mainly of
singly excited configuration (H-1 $\rightarrow$ L) and its charge
conjugated (c.c.) counterpart (H $\rightarrow$ L+1), where H$\equiv$HOMO and
L$\equiv$LUMO. 
Feature II located near 2.7 eV has a smaller intensity compared to
I, and corresponds to two closely-spaced $^3A_g$ states also consisting
predominantly of single excitations (H $\rightarrow$ L+3) + c.c, and 
(H-1 $\rightarrow$ L+2) + c.c. The intensity of the next feature (III) 
is much stronger compared to that of II, and is composed of states with main
contributions from {\em double excitations}. The dominant peak of III, located
near 3.9 eV, corresponds to a $^3A_g$ state composed predominantly of 
double excitation (H-1 $\rightarrow$ L ; H $\rightarrow$  L+1).
Feature IV has weaker intensity compared to III, and its main peak
occurs close to 5.1 eV. Similar to III, this feature also corresponds
to states composed of doubly excited configurations. The state
constituting the main peak of IV is predominantly composed of
 (H $\rightarrow$ L+2 ; H-2 $\rightarrow$ L), and other double 
excitations. In the infinite polymer limit, we speculate that the
single-excitation-based features I and II will merge to form the
first absorption band, while the double-excitation-based features
III and IV would merge to form the second band. Thus, for infinite
PPV, we expect the triplet PA spectrum to be similar in appearance 
to its singlet counterpart, which also exhibits two major
bands PA1 and PA2.~\cite{vardeny} 

Experimental measurement of triplet PA peaks beyond $m^3A_g$ (peak I) is not
an easy task.~\cite{val}
  However, if it is possible to conquer the associated practical difficulties,
we believe that the measurements of the high-energy region of the triplet PA
will shed light on some important excited states of PPV, thereby further
clarifying the role of electron correlation effects in this substance.

\section{Conclusions and Future Directions}
\label{conclusion}
In conclusion, very large-scale correlated calculations using
the PPP Hamiltonian were performed on the oligo-PPVs, and their 
triplet PA spectrum extending well into the high-energy region was computed.
The computed spectrum predicts rich structure beyond $m^3A_g$, 
consisting of peaks with alternating
high and low intensities. The peaks were shown to result from $^3A_g$-type
states composed predominantly of configurations which, besides 
the spin excitation, also
exhibit either single or double charge excitations with
respect to the ground state.  Therefore, it will be interesting to
probe the higher-energy triplet PA spectrum both of oligomers, 
as well as of infinite PPV experimentally, as it will lead 
to further insights
into the low-lying excited states of this technologically important
material. Analogous to the singlet manifold, it will then also be of
interest to explore the existence of triplet biexcitons. 

Although the electron-correlation effects were included to a large accuracy
in the present calculations, yet, because of the use of a rigid-lattice
model, the influence of electron-lattice coupling was completely ignored. 
However, recently Barford et al.~\cite{bursill} have demonstrated that
the contribution of electron-lattice relaxation can be important on the
low-lying correlated triplet states. They demonstrated that for polyenes
$1^3B_u$ state can relax by a few tenths of an eV thereby increasing the
corresponding  $E(m^3A_g)- E(1^1B_u)$ energy gap.~\cite{bursill} Therefore,
it will be of considerable interest to perform a similar calculation for
oligo-PPVs, and compute the contribution of electron-phonon coupling on
the triplet excitation energies of these materials.
These aspects
as well as phosphorescence and triplet 
electroluminescence in conjugated polymers will be studied theoretically in 
future publications.

\acknowledgements
The author is grateful to S. Mazumdar for a critical reading of the
manuscript, and for numerous suggestions for improvement. 
He also gratefully acknowledges the award of a guest scientist position by
Max-Planck-Institut f\"ur Physik Komplexer Systeme (MPIPKS), Dresden, Germany,
where part of this work was performed. 
These calculations were performed on the Alpha workstations of Physics
Deparment, IIT Bombay, and MPIPKS.

\clearpage
\newpage
\clearpage
\newpage
\begin{table}
\caption{For different states of various oligomers, the number of reference
configurations ($N_{ref}$) and the total number of configurations ($N_{total}$) 
involved in the MRSDCI (or QCI, where indicated) calculations.}
\begin{tabular}{lrlrlrlrl} \hline
Oligomer & \multicolumn{2}{c}{$1^1A_g$} &  \multicolumn{2}{c}{$1^1B_u$} &
\multicolumn{2}{c}{$1^3B_u$} & \multicolumn{2}{c}{$m^3A_g$} \\
     &  $N_{ref}$ & $N_{total}$ &  $N_{ref}$ & $N_{total}$ &  $N_{ref}$ & $N_{total}$ 
&  $N_{ref}$ & $N_{total}$  \\ \hline 
PPV-2  & 1$^a$ & 49495 & 1$^a$ & 82423 & 1$^a$ & 144543$^a$ & 72 &  112193 \\
PPV-3  & 1$^a$ & 2003907 & 1$^a$ & 3416371 & 65 & 1212746  & 65 & 1321885 \\ 
PPV-4  & 35 &1627923 & 36 & 1503239 & 35 & 2136547 & 35  & 2766111 \\
PPV-5  & 25 & 3257930 & 25 &  2537197 & 25 & 3217908 & 20  & 3314575 \\   
\hline
\end{tabular}
$^a$ QCI method
\label{tab-nref}
\end{table}
\begin{table}
\caption{Comparison of our results with the standard parameters (This work-1)
and screened parameters (This work-2) with those of other authors and 
experiments.  All the energies are in eV.}
\begin{tabular}{llllll}
\hline
                   & Work     & PPV-2   & PPV-3 & PPV-4 & PPV-5 \\ \hline
$1^1A_g$ -$1^1B_u$ & This work-1 &   4.48  & 4.11  & 3.93  & 3.83  \\
                   & This work-2 &   4.34  & 3.91  & 3.65  & 3.46 \\
                    & Beljonne et al.~\cite{beljonne} 
                                &   4.48  & 3.88  & 3.53  & 3.47  \\
                    & Lavrentiev et al.~\cite{barford}
                                 & 4.17   & 3.52 & 3.18 & 2.99 \\
                   & Exp.   &   3.71$^a$ & 3.43$^b$  & 3.20$^b$  & 3.07$^b$  \\     
$1^1A_g$ -$1^3B_u$ & This work-1 & 2.66    & 2.64  &   2.63 &   2.62 \\
                   & This work-2 & 2.20    & 2.19  &   2.22 &   2.21 \\    
                   & Beljonne et al.~\cite{beljonne}
                               & 2.73    & 2.52  &  2.44  &   2.63 \\
                    & Lavrentiev et al.~\cite{barford}
                                 & 2.65   & 2.16 & 1.95 & 1.84 \\
$1^3B_u$ -$m^3A_g$ & This work-1 & 3.74    & 3.16  &  2.92  &  2.90   \\
                   & This work-2 & 3.25    & 2.55  &  2.14      &  1.93    \\ 
                   & Beljonne et al.~\cite{beljonne}
                               & 4.15     & 3.26  &  2.75  &  2.46  \\
                   & Lavrentiev et al.~\cite{barford}
                                 & 3.63   & 3.04 & 2.61 & 2.34 \\
                   & Exp.      & 3.29$^c$ & 2.30$^d$  &  1.95$^d$  & 1.80$^d$ \\
\end{tabular}
\label{tab-comp}
$^a$ Ref.~\cite{dyck}    \hspace*{30pt}     $^c$ Ref.~\cite{herks} \\
$^b$ Ref.~\cite{gelinck} \hspace*{30pt}  $^d$ Ref.~\cite{cerullo} 
\end{table}

\end{document}